\documentclass{aa}  
\usepackage{graphicx}
\usepackage{txfonts}
\usepackage{natbib}
\begin{document}
   \title{CO and CH$_3$OH observations of the BHR71 outflows with APEX}


   \author{B. Parise
          \and
          A. Belloche
          \and 
          S. Leurini
          \and
          P. Schilke
          \and
          F. Wyrowski
          \and 
          R. G\"usten
          }

   \offprints{B. Parise}

   \institute{Max Planck Institut f\"ur Radioastronomie, Auf dem H\"ugel 69, 53121 Bonn, Germany\\
              \email{[bparise,belloche,sleurini,schilke,wyrowski,rguesten]@mpifr-bonn.mpg.de}
             }

   \date{Received September 15, 1996; accepted March 16, 1997}

 
  \abstract
  {Highly-collimated outflows are believed to be the earliest stage in outflow evolution, so their study is essential for understanding the processes driving outflows. The BHR71 Bok globule is known to harbour such a highly-collimated outflow, which is powered by a protostar belonging to a protobinary system.   }
   {We aimed at investigating the interaction of collimated outflows with the ambient molecular cloud by using molecular tracers. }
  {We mapped the BHR71 highly-collimated outflow in CO(3-2) with the APEX telescope, and observed several bright points of the outflow in the molecular transitions CO(4-3), $^{13}$CO(3-2), C$^{18}$O(3-2), and CH$_3$OH(7-6). We use an LVG code to characterise the temperature enhancements in these regions.}
  {In our CO(3-2) map, the second outflow driven by IRS2, which is the second source of the binary system, is completely revealed and shown to be bipolar. We also measure temperature enhancements in the lobes. The CO and methanol LVG modelling points to temperatures between 30 and 50\,K in the IRS1 outflow, while the IRS2 outflow seems to be warmer (up to 300\,K).}
   {}
   

   \maketitle
%

\section{Introduction}

Low-mass stars form in molecular clouds from the collapse of dense cores. In the earliest stage,
known as Class 0, the newly-born protostar is deeply embedded in a thick envelope of gas and dust from which it will accrete most of its mass, and it drives powerful outflows. However, the mechanisms driving bipolar outflows, as well as the interaction of the outflow with the surrounding material, are still not understood well.
It is commonly believed that highly-collimated outflows represent the earliest stage in outflow evolution,
making their observation central to getting new insights into the mechanisms driving outflows. 

The BHR71 isolated Bok globule, located at $\sim$\,200 pc, harbours one of the best examples of such a highly-collimated outflow. This outflow, which is almost lying in the plane of the sky, was first discovered and mapped in $^{12}$CO(1-0) and   $^{13}$CO(1-0) by \citet{Bourke97}. It appears to be driven by the very young stellar object IRAS11590$-$6452, classified as a Class 0/I protostar by \citet{Froebrich05}, with a total luminosity of $\sim$ 9 L$_{\odot}$. The CO emission was observed to peak towards the outflow lobes, implying that the ambient gas is heated by interaction with the outflow. 

Subsequent observations by \citet{Garay98} aimed at constraining the magnitude of the temperature enhancement in the lobes, by mapping several transitions of CS, SiO, CH$_3$OH, and HCO$^+$ with the SEST telescope. Assuming n\,=\,10$^5$\,cm$^{-3}$, the CH$_3$OH (2-1) and (3-2) observations were found to be consistent with kinetic temperatures in the 50-100\,K range.

ISOCAM observations of BHR71 showed that the IRAS source is in fact a multiple system \citep{Myers98}. Two protostellar objects IRS1 and IRS2 are present in BHR71, and the large collimated outflow is driven by IRS1 \citep{Bourke01}. Combination of CO(2-1), NIR, and cm observations shows  evidence of the presence of the blue lobe of a second, fainter outflow driven by IRS2 \citep{Bourke01}.

In this Letter, we present CO(3-2) observations that fully unveil the presence of this second outflow, and characterise the temperature enhancement towards bright knots of the larger outflow by observations of CO(4-3), $^{13}$CO(3-2), $^{18}$CO(3-2), and CH$_3$OH(7-6).


\section{Observations}

   \begin{figure*}
   \centering
   \begin{tabular}{lll}
   \includegraphics [width=0.40\textwidth,angle=-90]{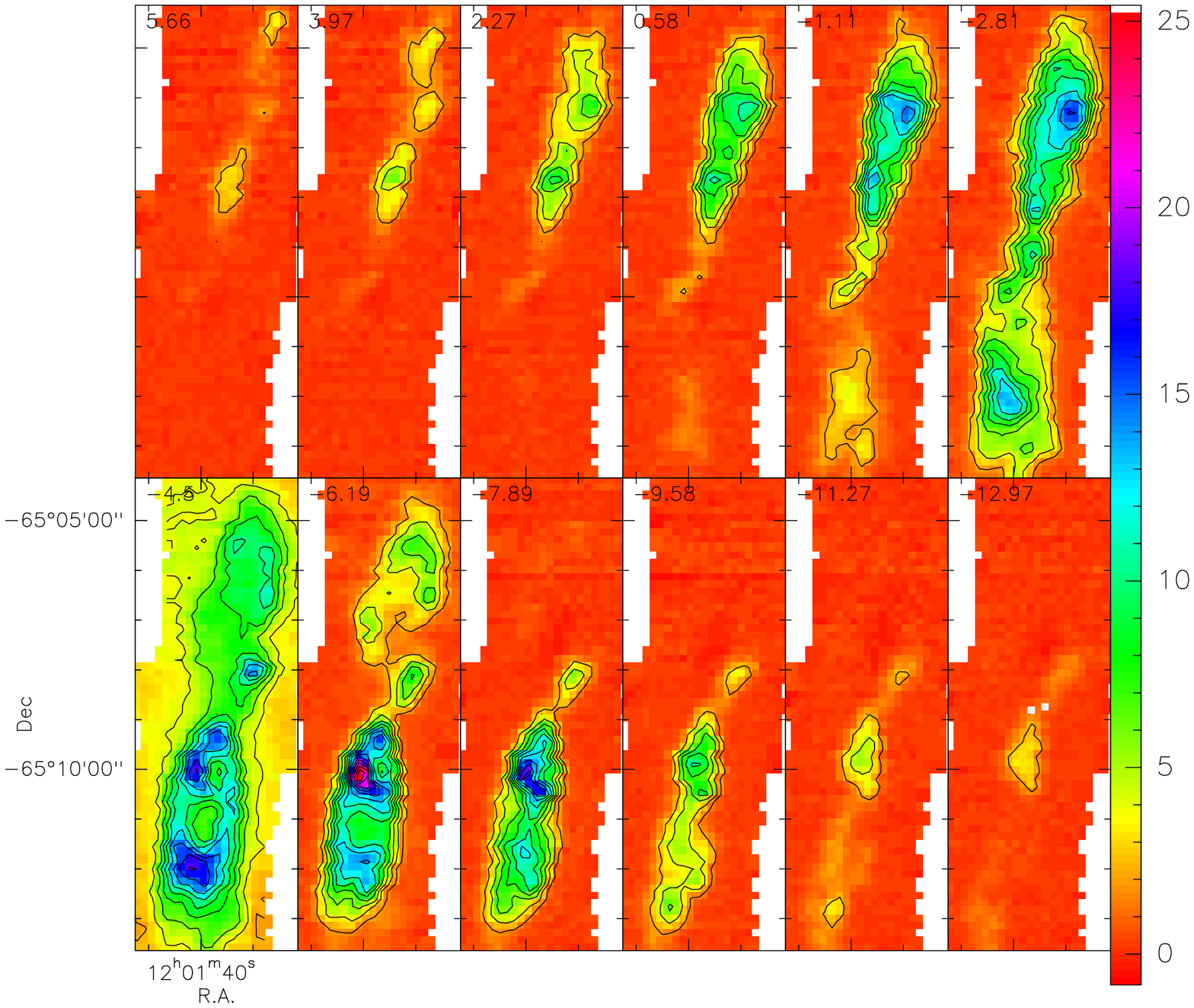} & 
   \includegraphics[width=0.40\textwidth,angle=-90]{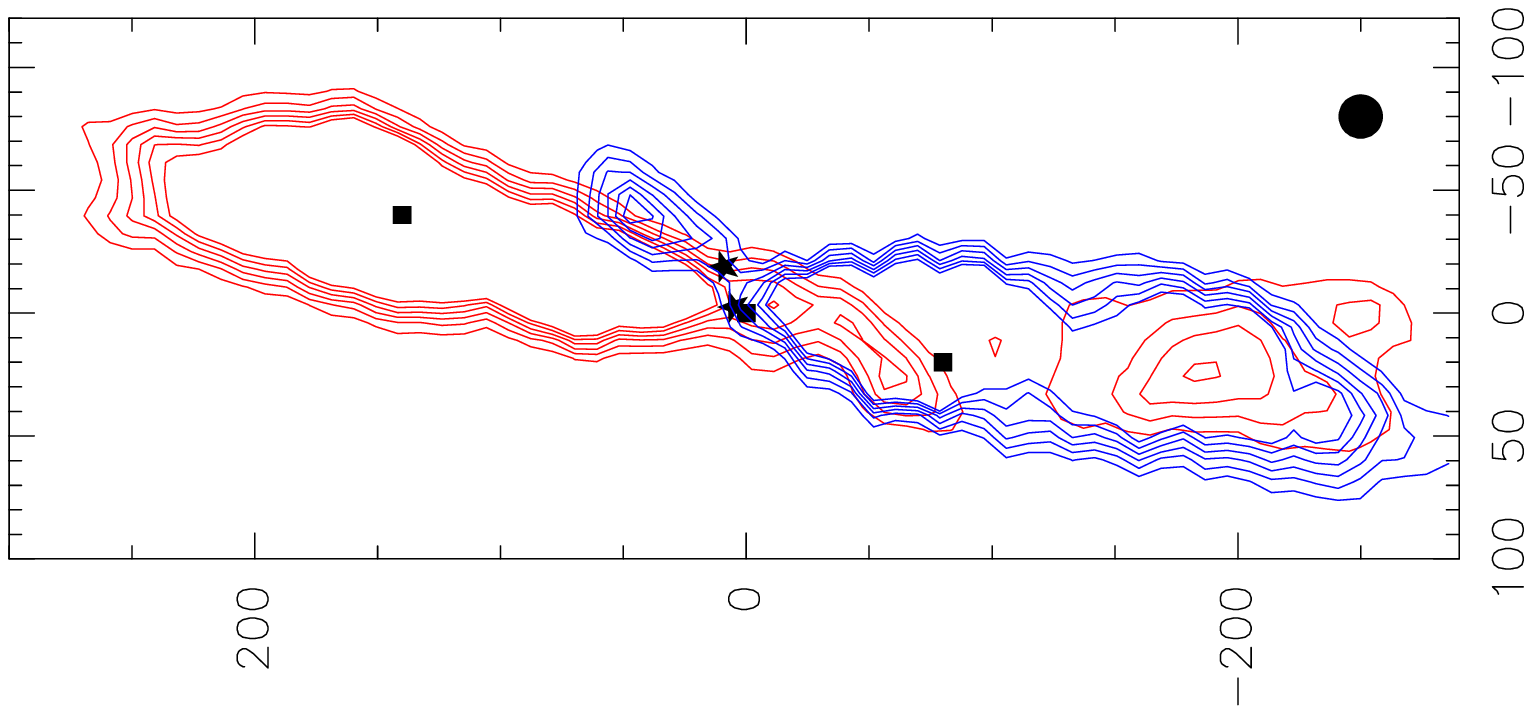} &
   \includegraphics [width=0.41\textwidth,angle=-90]{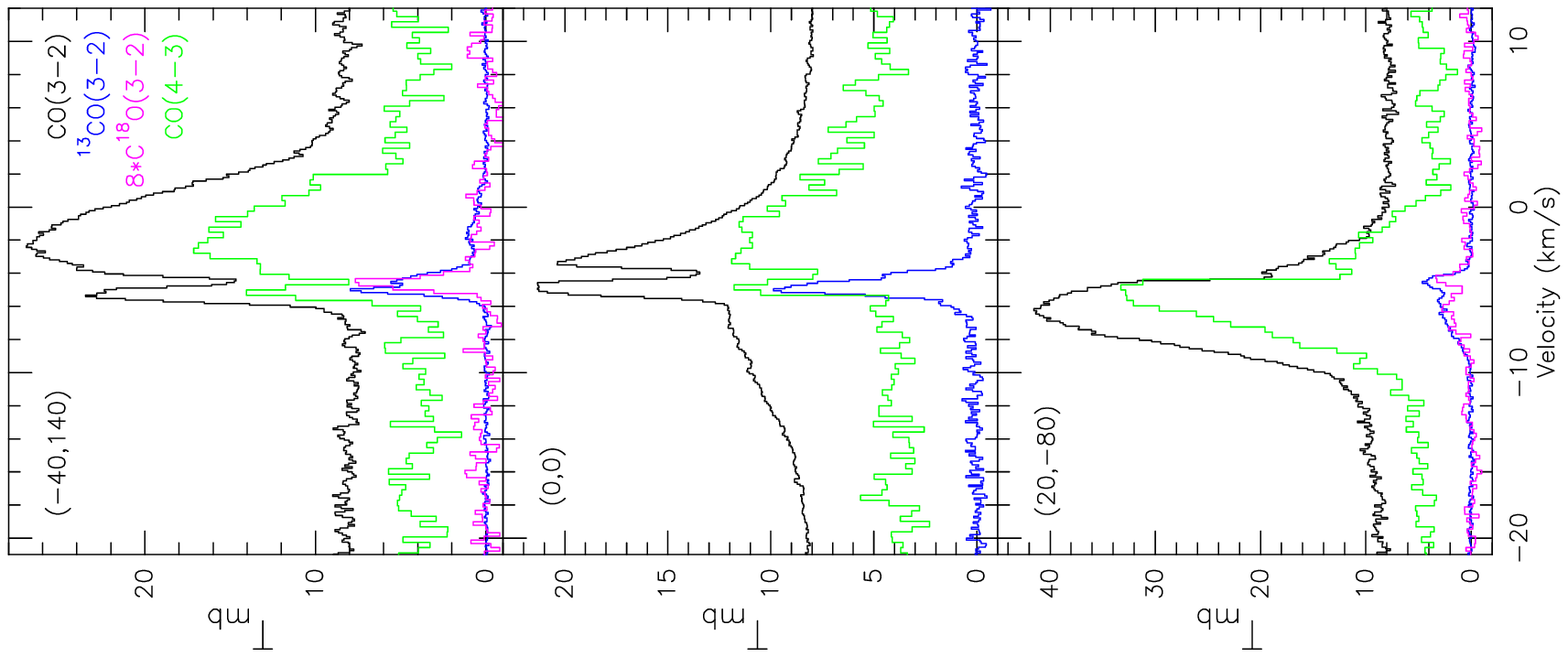}\\
    \vspace{-1.2cm}\\
   (a) & \hspace{-0.3cm}(b) & (c) \\
   \end{tabular}
   \vspace{0.4cm}
      \caption{(a) CO(3-2) channel maps (centred on $\alpha$(2000)\,=\,12$^{\rm h}$01$^{\rm m}$37$^{\rm s}$, $\delta$(2000)\,=\,$-$65$^{\circ}$08$'$53.5$''$) on a $\int$ T${\rm _A^*}$\,dv scale (the v$_{\rm lsr}$ of the cloud is \hbox{-4.5}\,km\,s$^{-1}$). (b) Map of the integrated CO(3-2) emission in the -14$<$v$<$-6 (blue) and -3$<$v$<$5\,km\,s$^{-1}$ (red) intervals, enlightening the two bipolar outflows. Contours are from 7.5 to 30 in steps of 4 K\,km\,s$^{-1}$ . Contours for higher flux (in the north-south outflow) have not been plotted in order to show the fainter outflow better. The two black stars show the position of the two protostellar sources. The black circle in the lower right corner shows the beam size. (c) CO(3-2), $^{13}$CO(3-2), C$^{18}$O(3-2), and CO(4-3) observations on the three positions marked as black squares in (b). For clarity, CO(3-2) and CO(4-3) lines have been shifted by 8 and 4, respectively, along the $y$-axis.}
         \label{co}
   \end{figure*}

Observations towards BHR71 were carried out in September and November 2005 with the Atacama Pathfinder EXperiment (APEX\footnote{This publication is based on data acquired with the Atacama Pathfinder Experiment (APEX). APEX is a collaboration between the Max-Planck-Institut f\"ur Radioastronomie, the European Southern Observatory, and the Onsala Space Observatory.}, G\"usten et al. 2006, {\it this volume}), using the facility's own APEX2a receiver (Risacher et al. 2006, {\it this volume}) and the MPIfR First Light APEX Submillimeter Heterodyne receiver
(FLASH, Heyminck et al. 2006, {\it this volume}), along with the MPIfR Fast Fourier Transform Spectrometer backend 
(FFTS, Klein et al. 2006, {\it this volume}).
The beam of the telescope is 17.3$''$ at 345\,GHz and 13.3$''$ at 460\,GHz (G\"usten et al. 2006, {\it this volume}).
Focus was checked at the beginning of each run and after sunrise on Saturn. Pointing was checked on Saturn, 07454-7112 or IRAS15194-5115. The pointing accuracy was found to be of the order of 4$''$ rms.

We mapped the large-scale outflow in the CO(3-2) transition using the on-the-fly observing mode. 
Several other transitions were observed towards several bright points of the outflow (CO(4-3), $^{13}$CO(3-2), C$^{18}$O(3-2), and the CH$_3$OH 7$_k\to$6$_k$ band). The double-side-band system temperatures ranged from 100\,K to 200\,K for all observations, except for CO(4-3) where they were in the range 1000-3000\,K. We used 8192 channels of the FFTS over the 1\,GHz bandwidth, yielding an effective resolution of  about 240 kHz.
The observations were performed in position-switching mode using the APECS software (Muders et al. 2006, {\it this volume}). The data were reduced with the CLASS software (see http://www.iram.fr/IRAMFR/GILDAS). 
Antenna temperatures were converted to main-beam temperatures using forward and beam efficiencies, respectively, of 0.97 and 0.73 at 345\,GHz and of 0.95 and 0.61 at 464\,GHz (G\"usten et al. 2006, {\it this volume}).


\section{Analysis}

\subsection{Outflow morphology}

The channel map of the CO(3-2) transition (Fig. \ref{co}a) displays the beautifully collimated outflow discovered by \citet{Bourke97}. Its PA is $\sim$165$^{\circ}$-170$^{\circ}$, and its projected collimation factor is 4 (as computed on the half-power contour). The cloud emission arises between -6 and -3\,km\,s$^{-1}$. Note that the two lobes are well separated spatially, and that there is a small amount of red emission in the blue lobe and vice versa. These two facts confirm that the inclination of the outflow with respect to the plane of the sky is close to its semi-opening angle ($\sim$15$^{\circ}$), as already noted by \citet{Bourke97}. The channel map also shows hints of a second fainter outflow that is a bit tilted relative to the bigger one, with the blue and red lobes inverted. This is clearer in Fig. \ref{co}b, where contours from emission for velocities in the [-14,-6] and [-3,5] km\,s$^{-1}$ range have been plotted. The second outflow is centred on the second source of the binary, IRS2. Its PA is between $-$35$^{\circ}$ and $-$30$^{\circ}$ and its projected semi-opening angle $\sim$\,25$^{\circ}$. Its projected collimation factor is 2. No red emission seems associated with the blue lobe, and the blue and red lobes are well separated spatially, so the inclination of this outflow is between 25$^{\circ}$ and 65$^{\circ}$ \citep{Cabrit86}. 

\subsection{Temperature enhancement in the lobes}

\subsubsection{CO optical depth}

C$^{18}$O was observed towards the two bright positions of the outflow at offsets (20$''$,\,-80$''$) and (-40$''$,\,140$''$). 
We assume isotopic ratios $^{12}$C/$^{13}$C\,=\,69$\pm$6 and $^{16}$O/$^{18}$O\,=\,557$\pm$30 \citep{Wilson99}. The expected $^{13}$CO/C$^{18}$O abundance ratio is then 8.1{\small $^{+1.5}_{-1.3}$}. As shown in Fig. \ref{co}c, the ratio between the antenna temperature of the two isotopologues is found to lie in the 8-10 range, so we conclude that the 3-2 lines of both isotopologues are optically thin. 

From the $^{13}$CO(3-2), we can estimate the optical depth in the CO(3-2) transition. We computed the opacity in various velocity channels (see Table \ref{lvg}). The highest opacity is measured on the (0$''$,\,0$''$) position. Towards the two lobe positions, the opacity is lower: $\sim$\,5 for the (20$''$,\,-80$''$) offset position and $\sim$\,3 for (-40$''$,\,140$''$).

\subsubsection{CO excitation}
\label{colvg} 
\vspace{-0.3cm}

\begin{table}[h!]
\caption{CO LVG modelling.}
\label{lvg}
\begin{footnotesize}
\begin{tabular}{ccccc}
\hline
\hline
\noalign{\smallskip}
Vel channel  &   $\tau_{32}$    &    N$_{\rm CO}$   &    T$_{\rm mb, (4-3)}$/T$_{\rm mb, (3-2)}$  &  ${\rm T_k}$      \\ 
 (km\,s$^{-1}$) &  &  (cm$^{-2}$)  & & (K)\\
\noalign{\smallskip}
\hline
\noalign{\smallskip}
\multicolumn{5}{c}{Offset (-40$''$,\,140$''$)  -  Northern lobe (big outflow)} \\
\noalign{\smallskip}
\hline
\noalign{\smallskip}
$[$-3,\,0$]$  &  3.2  &  1.3$\times$10$^{17}$  &  0.67  &  $>$ 35 \\
$[$0,\,2$]$    &  1.4  &  2.9$\times$10$^{16}$   & 0.65 & $>$ 20 \\
\noalign{\smallskip}
\hline
\noalign{\smallskip}
\multicolumn{5}{c}{Offset (0$''$,\,0$''$)  } \\
\noalign{\smallskip}
\hline
\noalign{\smallskip}
\multicolumn{5}{c}{Blue component (IRS1 outflow)} \\
$[$-7,\,-6$]$ & 13 & 3.8$\times$10$^{16}$ & 0.20 & $>25$\\
\multicolumn{5}{c}{Red component (both outflows)} \\
$[$-3,\,-2$]$ & $<$3.5  &  $<$2.4$\times$10$^{16}$ & 0.79 & $>$20$^1$\\
           &                  &                                                &        & $>$20$^2$\\
$[$-2,\,-1$]$ & $<$5.5  & $<$2.5$\times$10$^{16}$ & 1.2 & $>$80$^1$\\
           &                  &                                                &        & $>$60$^2$\\
$[$-1,\,0$]$ & $<$8.1  & $<$2.5$\times$10$^{16}$ & 1.5 & $>$190$^1$\\
           &                  &                                                &        & $>$130$^2$\\
\noalign{\smallskip}
\hline
\noalign{\smallskip}           
\multicolumn{5}{c}{Offset (20$''$,\,-80$''$)  -  Southern lobe} \\
\noalign{\smallskip}
\hline
\noalign{\smallskip}
\multicolumn{5}{c}{Blue component (IRS1 outflow) } \\
$[$-9,\,-6$]$ & 4.8 & 2.8$\times$10$^{17}$  & 0.54 & $>50$ \\
\multicolumn{5}{c}{Red component (emission dominated by IRS2 outflow) }\\
$[$-3,\,0$]$ & $<$3.0 & $<$1.7$\times$10$^{16}$ & 1.9$^*$ & $>320^3$ \\
                     &               &                                                 &                & $>300^4$\\
\noalign{\smallskip}
\hline
\noalign{\smallskip}
\end{tabular}
$^*$ 1.7\,$<$\,T$_{\rm mb, (4,3)}$/T$_{\rm mb, (3,2)}$\,$<$\,2.1 (1$\sigma$).  \\
$^1$ Assuming N$_{\rm CO}$/$\Delta$v=3$\times$10$^{16}$ cm$^{-2}$\,km$^{-1}$\,s.\\
$^2$ Assuming N$_{\rm CO}$/$\Delta$v=3$\times$10$^{15}$ cm$^{-2}$\,km$^{-1}$\,s.\\
$^3$ Assuming N$_{\rm CO}$/$\Delta$v=6$\times$10$^{15}$ cm$^{-2}$\,km$^{-1}$\,s.\\
$^4$ Assuming N$_{\rm CO}$/$\Delta$v=6$\times$10$^{14}$ cm$^{-2}$\,km$^{-1}$\,s.
\vspace{-0.4cm}
\end{footnotesize}
\end{table}

\begin{figure}[!h]
\includegraphics[width=0.48\textwidth,angle=-90]{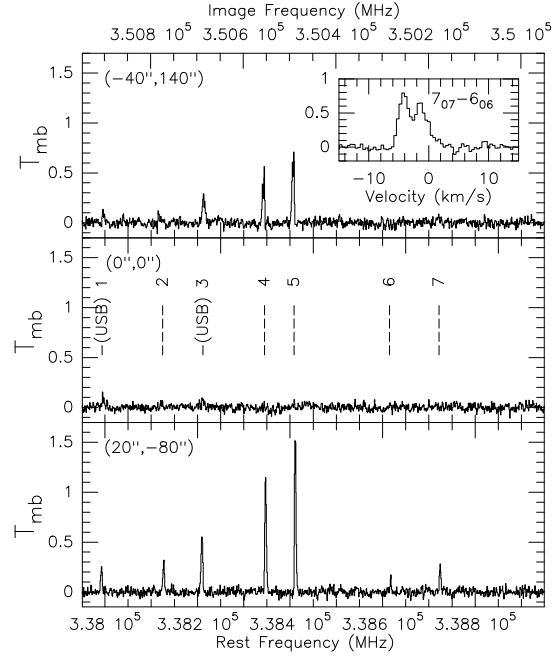}
\caption{Methanol spectra observed towards the 3 positions shown in Fig.\,1b,\,c. The vertical dashed numbered lines in the middle panel refer to the following transitions: (1) 1$_{1,1}$-0$_{0,0}$ [USB], (2) 7$_{0,7}$-6$_{0,6}$, (3) 4$_{0,4}$-3$_{-1,3}$ [USB], (4) 7$_{-1,7}$-6$_{-1,6}$, (5) 7$_{0,7}$-6$_{0,6}$, (6) 7$_{+1,6}$-6$_{+1,5}$, (7) 7$_{2,5}$-6$_{2,4}$, \& 7$_{-2,6}$-6$_{-2,5}$. The inset shows the double-peak structure of the 7$_{0,7}$-6$_{0,6}$ line towards the northern position.}
\label{ch3oh}
\vspace{-0.3cm}
\end{figure}

\begin{figure*}[!ht]
\begin{center}
\includegraphics[width=0.36\textwidth,angle=90]{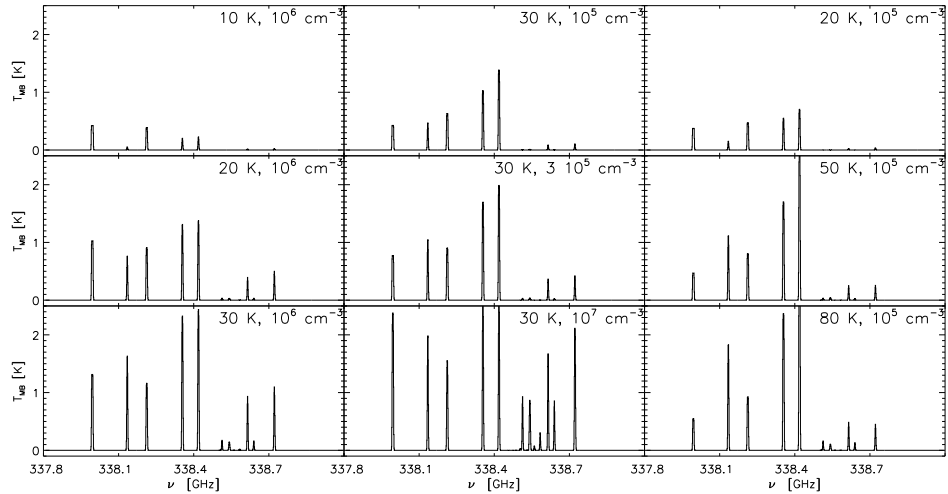}
\end{center}
\vspace{-0.4cm}
\caption{Methanol synthetic spectra. A CH$_3$OH column density of 10$^{16}$ cm$^{-2}$ and a source size of 7$''$ were assumed. }
\label{ch3oh_syn}
\vspace{-0.4cm}
\end{figure*}


Using a standard spherically symmetric large-velocity-gradient (LVG) code, we can derive the kinetic temperature in the outflow lobes from our CO(3-2), $^{13}$CO(3-2), and CO(4-3) observations. The inputs of the code are the CO column density, the kinetic temperature, and the density. 
The CO column density can be derived from the CO(3-2) optical depth as shown by \citet{Choi93}. 
In each velocity channel $i$, the column density is given by:
$$ {\rm N_i^{CO} {\it (cm^{-2})} = 1.1 \times 10^{15} ~\frac{T_{mb,32} \Delta v~  {\it (K\,km\,s^{-1})}}{D(n,T_k)} ~\frac{\tau_{32}}{(1- exp(-\tau_{32}))} }$$
where D(n,T$_k$) was shown to be equal to 1.5 within a factor of 2 in the whole range $10<T_k<200$K and $10^4<n<10^6$ cm$^{-3}$.  
All the computations are done assuming that the beam-filling factor is one for each transition. However, the CO(4-3) emission may be less extended than the CO(3-2) emission, so the derived $T_k$ and $n$ are lower limits. 
Table \ref{lvg} summarises the results of this modelling. The constraint on the density is found to be not significant (n\,$>$\,10$^3$\,cm$^{-3}$), and was thus not reported.

Modelling of the CO emission shows temperature enhancements from 20 to 50\,K towards the IRS1  outflow, which is consistent with the temperature derived from methanol analysis (see Sect. 3.2.3).
It should be noted that both the (0$''$,\,0$''$) and the (20$''$,\,-80$''$) offset positions comprise 
emission from the small outflow. In particular, the red emission at the (20$''$,\,-80$''$) position is dominated by this second outflow. The derived temperature is very high, as the CO(4-3) emission is stronger than the CO(3-2). The outflow powered by IRS 2 thus appears to be warmer than the IRS1 outflow. This could be caused by the observation targets including the tips of the IRS2 outflow cavity, thereby tracing the present shocks, while toward IRS1, the observed position corresponds to a position in the centre of the flow, which could trace colder postshock gas.
Note that the HH321B shock \citep{Corporon97,Bourke01} is encompassed in the beam of our observations at the (20$''$,-80$''$) position, and high-resolution IR observations would be very useful for  determining if this HH object is indeed associated with the IRS2 outflow, as suggested by the association of the HH320 objects with its northern lobe \citep{Bourke01}.

\subsubsection{Methanol statistical equilibrium calculations}


The CH$_3$OH $7_k \to 6_k$ band at 338\,GHz was observed towards the core
position and two bright positions of the main outflow with the
APEX-2a receiver tuned in the lower side band (Fig. \ref{ch3oh}). This frequency setup
allows for simultaneous detection in the upper sideband of the 
$4_{04}$-3$_{-13}$ line at
350.688\,GHz and of the 
$1_{11}$-$0_{00}$ transition at 350.905\,GHz,
which have low excitation energies (${\rm T_{low}<20}$\,K) compared to the 
$7_k \to 6_k$ band (${\rm T_{low}>44}$\,K). The three positions differ
significantly in the methanol emission. Towards the central position only weak
emission is detected from the $4_{04}$-3$_{-13}$, the $1_{11}$-$0_{00}$, and maybe the 
$7_{07}$-6$_{06}$ 
transitions. In contrast, lines from the
$7_k \to 6_k$ are also observed towards the two lobe positions, the lower energy transitions toward the northern lobe, and moderately 
higher energy transitions towards the south position. In the (20$''$,\,$-$80$''$) offset position,
the methanol emission is blueshifted, and thus associated with the IRS1 outflow. 
In the ($-$40$''$,\,140$''$) offset position, the lines are double-peaked, with a redshifted peak and a contribution at the cloud velocity (see inset in Fig. \ref{ch3oh}), which might be associated with emission from the preshock material, where methanol is released from the grains, but not accelerated yet.

\citet{Leurini04} concluded that low-excitation CH$_3$OH 
transitions often trace the density of the gas, while highly excited
lines are also sensitive to the kinetic temperature. Methanol can therefore be 
used as a tracer of both temperature and density. To determine the physical parameters we used a spherically symmetric large velocity gradient
statistical equilibrium code, with the cosmic background as the only
radiation field.  Line ratios between different transitions can be used to constrain the temperature and density. We assumed that all lines are emitted in the same gas; {\it i.e.} we suppose that all lines have the same beam-filling factor. This assumption is likely to be correct, as all the detected transitions have similar lower energy ($<$\,75\,K, and even between 44 and 75\,K, if we only focus on the $7_k\to 6_k$ band). 

Fig. \ref{ch3oh_syn} presents synthetic double-side-band spectra for different temperature and density conditions, to be compared with our observations. We conclude that, towards the (20$''$,\,-80$''$) position, CH$_3$OH comes from a relatively warm (30 $<$  T $<$ 50\,K) and dense (10$^5$ $<$ n(H$_2$) $<$ 3$\times$10$^5\,{\rm{cm}^{-3}}$) gas. The temperature enhancement that we derive is a bit lower than the range suggested by \citet[][ between 50 and 100\,K, as deduced from analysis of lower energy lines]{Garay98}.  The transitions they used span a narrower range of excitation conditions than the presently observed ones, so they are less sensitive to temperature, as shown by \citet{Leurini04}.
The detection of only the 
$7_{-17}$-6$_{-16}$ and $7_{07}$-6$_{06}$ lines 
towards the red lobe implies that the gas temperature must be around 20 to 30\,K and not denser than a few $10^5~{\rm{cm}^{-3}}$. 
Finally, the non detection of the $7_k \to 6_k$ lines towards the core position infers a temperature $\sim$\,10\,K (computation for N$_{\rm CH_3OH}$=10$^{16}$cm$^{-2}$).\footnote{For lower column densities, more appropriate for the core position, the detection of only the 1$_{11}$-0$_{00}$ transition points even more clearly to low temperatures.}

\section{Conclusion}

From CO and methanol observations, we have constrained the density and temperature enhancement in the lobes of the highly-collimated outflow of BHR71. Excitation analysis points to temperatures between 30 and 50\,K, while the density is around 10$^5$\,cm$^{-3}$.
Our observations have allowed us to isolate and determine the structure of the second outflow in BHR71 driven by IRS2. From CO excitation analysis, this outflow seems to have very high temperatures.
Further observations to better constrain the temperature enhancement of this second outflow (e.g. the methanol $7_k \to6_k$ band towards the blue lobe) would be very valuable. 

\begin{footnotesize}
\begin{acknowledgements}
BP is grateful to the \emph{Alexander von Humboldt Foundation} for a Humboldt Research Fellowship.
\end{acknowledgements}

\bibliography{/Users/bparise/These/Manuscrit/biblio}
\bibliographystyle{aa}
\end{footnotesize}
\end{document}